\documentclass[twoside,twocolumn,9pt]{article}
\usepackage[utf8]{inputenc}
\usepackage{extsizes}
\usepackage[super,sort&compress,comma]{natbib}
\usepackage[version=3]{mhchem}
\usepackage[left=1.5cm, right=1.5cm, top=1.785cm, bottom=2.0cm]{geometry}
\usepackage{balance}
\usepackage{widetext}
\usepackage{times,mathptmx}
\usepackage{sectsty}
\usepackage{graphicx}
\usepackage{lastpage}
\usepackage[format=plain,justification=raggedright,singlelinecheck=false,font={stretch=1.125,small,sf},labelfont=bf,labelsep=space]{caption}
\usepackage{float}
\usepackage{fancyhdr}
\usepackage{fnpos}
\usepackage[english]{babel}
\usepackage{array}
\usepackage{droidsans}
\usepackage{charter}
\usepackage[T1]{fontenc}
\usepackage[usenames,dvipsnames]{xcolor}
\usepackage{setspace}
\usepackage[compact]{titlesec}

\usepackage[
	colorlinks,
    linkcolor=black,
    citecolor=black,
    urlcolor=black
]{hyperref}
\usepackage{microtype}
\usepackage{bibentry}
\usepackage{fdsymbol}
\usepackage{xspace}

\definecolor{cream}{RGB}{222,217,201}

\begin{document}

\pagestyle{fancy}
\thispagestyle{plain}
\fancypagestyle{plain}{

\fancyhead[C]{\raisebox{5mm}{\colorbox[RGB]{146,164,183}{\parbox[l][11mm][c]{\textwidth}{\sffamily\Large\textcolor{white}{\bfseries\ \ \ FEATURE ARTICLE}}}}}
\fancyhead[L]{\hspace{0cm}\vspace{1.5cm}\sffamily\fontsize{24}{28}\selectfont\bfseries ChemComm}
\fancyhead[R]{%
\hspace{5cm}\raisebox{15mm}{\href{https://doi.org/10.1039/c7cc03306k}{\textcolor{blue}{%
Published as \emph{Chem. Commun.}, 2017, 53, 7211--7221}}}\hfill%
\hspace{0cm}\vspace{1.7cm}\includegraphics[height=55pt]{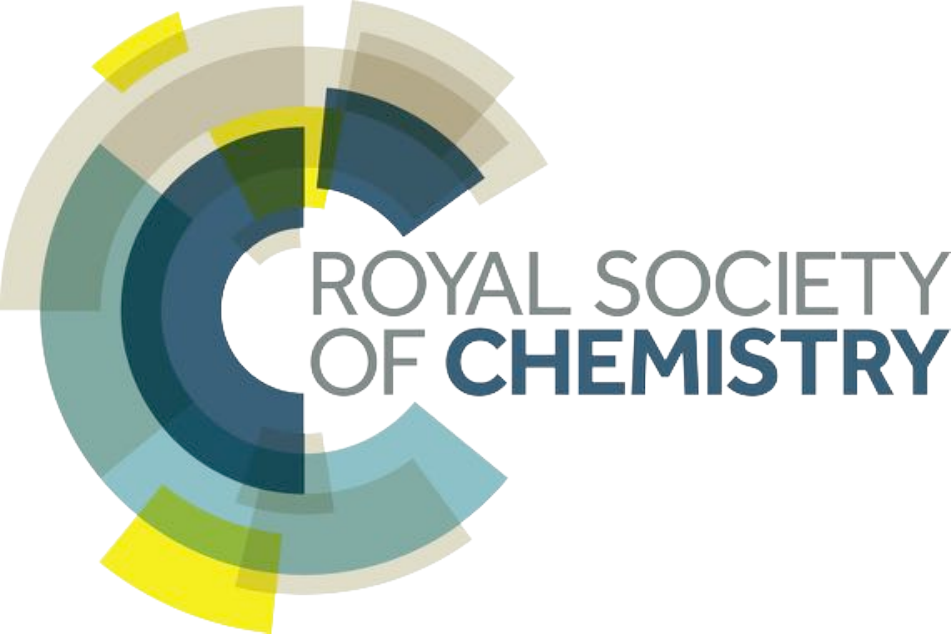}}
\renewcommand{\headrulewidth}{0pt}
}

\makeFNbottom
\makeatletter
\renewcommand\LARGE{\@setfontsize\LARGE{15pt}{17}}
\renewcommand\Large{\@setfontsize\Large{12pt}{14}}
\renewcommand\large{\@setfontsize\large{10pt}{12}}
\renewcommand\footnotesize{\@setfontsize\footnotesize{7pt}{10}}
\makeatother

\renewcommand{\thefootnote}{\fnsymbol{footnote}}
\renewcommand\footnoterule{\vspace*{1pt}%
\color{cream}\hrule width 3.5in height 0.4pt \color{black}\vspace*{5pt}}
\setcounter{secnumdepth}{5}

\makeatletter
\renewcommand\@biblabel[1]{#1}
\renewcommand\@makefntext[1]%
{\noindent\makebox[0pt][r]{\@thefnmark\,}#1}
\makeatother
\renewcommand{\figurename}{\small{Fig.}~}
\sectionfont{\sffamily\Large}
\subsectionfont{\normalsize}
\subsubsectionfont{\bf}
\setstretch{1.125} 
\setlength{\skip\footins}{0.8cm}
\setlength{\footnotesep}{0.25cm}
\setlength{\jot}{10pt}
\titlespacing*{\section}{0pt}{4pt}{4pt}
\titlespacing*{\subsection}{0pt}{15pt}{1pt}

\fancyfoot{}
\fancyfoot[LO,RE]{\vspace{-7.1pt}\sffamily\footnotesize\color{gray} This journal is {\textcopyright} The Royal Society of Chemistry 2017}
\fancyfoot[CO]{}
\fancyfoot[CE]{}
\fancyfoot[RO]{\footnotesize{\sffamily{\emph{Chem. Commun.}, 2017, [vol.], 1--\pageref{LastPage} ~\textbar  \hspace{2pt}\thepage}}}
\fancyfoot[LE]{\footnotesize{\sffamily{\thepage~\textbar\hspace{2mm}\emph{Chem. Commun.}, 2017, [vol.], 1--\pageref{LastPage}}}}
\fancyhead{}
\renewcommand{\headrulewidth}{0pt}
\renewcommand{\footrulewidth}{0pt}
\setlength{\arrayrulewidth}{1pt}
\setlength{\columnsep}{6.5mm}
\setlength\bibsep{1pt}

\makeatletter
\newlength{\figrulesep}\setlength{\figrulesep}{0.5\textfloatsep}
\newcommand{\topfigrule}{\vspace*{-1pt}%
\noindent{\color{cream}\rule[-\figrulesep]{\columnwidth}{1.5pt}} }
\newcommand{\botfigrule}{\vspace*{-2pt}%
\noindent{\color{cream}\rule[\figrulesep]{\columnwidth}{1.5pt}} }
\newcommand{\dblfigrule}{\vspace*{-1pt}%
\noindent{\color{cream}\rule[-\figrulesep]{\textwidth}{1.5pt}} }
\makeatother


\newcommand\revadd[1]{#1}
\newcommand*\revrem[1]{}

\newcommand\abinitio{\emph{ab initio}\xspace}

\twocolumn[\begin{@twocolumnfalse}\vspace{3cm}\sffamily
\begin{tabular}{m{4.2cm} p{13.5cm}}
\textbf{\scriptsize Cite this: DOI: 10.1039/xxxxxxxxxx}
& \noindent\LARGE{\textbf{Recent advances in the computational chemistry of \revadd{soft porous crystals}}} \\
\vspace{0.3cm} & \vspace{0.3cm} \\
& \noindent\large{\large Guillaume Fraux,$^{\text{a}}$ and Fran\c{c}ois-Xavier Coudert$^{\text{*a}}$} \\
\parbox{4cm}{\textbf{\footnotesize\\[1mm]
Received Date\\
Accepted Date\\[2mm]
DOI: 10.1039/xxxxxxxxxx\\[2mm]
www.rsc.org/
}}
& \noindent\normalsize{%
Here we highlight recent progress in the field of computational chemistry of nanoporous materials, focusing on methods and studies that address the extraordinary dynamic nature of these systems: high flexibility of their frameworks, large-scale structural changes upon external physical or chemical stimulation, presence of defects and disorder. The wide variety of behavior demonstrated in soft porous crystals, including the topical class of metal--organic frameworks, opens new challenges for computational chemistry methods at all scales.
} \\
\end{tabular}\end{@twocolumnfalse}\vspace{1.6cm}]

\renewcommand*\rmdefault{bch}\normalfont\upshape
\rmfamily
\section*{}
\vspace{-1cm}

\footnotetext{\textit{$^{a}$~Chimie ParisTech, PSL Research University, CNRS, Institut de Recherche de Chimie Paris, 75005 Paris, France; E-mail: \url{fx.coudert@chimie-paristech.fr}; Twitter: \href{https://twitter.com/fxcoudert}{@fxcoudert}; Website: \url{http://coudert.name/}}}


\section{Introduction}

Among the research community studying nanoporous materials, the past decade has seen a large focus of research effort on a novel subclass of materials that exhibit flexibility, i.e. large-scale changes in their structure which impact their physical and chemical properties.\cite{Horike2009} In contrast to inorganic nanoporous materials such as zeolites, there have been a large number of crystalline framework materials which are constructed from weaker interactions, e.g., \revadd{coordination} bonds, $\pi$--$\pi$ stacking, hydrogen bonds. This category includes the now ubiquitous metal--organic frameworks (MOFs),\cite{Furukawa2013} but also covalent organic frameworks (COFs)\cite{Feng2012} and supramolecular organic frameworks (SOFs).\cite{Holst2010} Based on the softer nature of their interactions, they may exhibit drastic changes in their internal structure --- and thus their properties --- upon stimulation by external physical or chemical stimuli. They can be affected by temperature, mechanical pressure, guest sorption, light, or magnetic field.\cite{Coudert2015}

Various terms have been used to describe these flexible materials, including \emph{dynamic}, \emph{smart} or \emph{multifunctional}.\cite{Schneemann2014} Here, we will use the terminology introduced by the Kitagawa group, referring to these highly flexible materials as \emph{soft porous crystals},\cite{Horike2009} or as \emph{stimuli-responsive materials}\cite{Coudert2015} for those that undergo large changes in their structure and properties. In addition to the conventional applications of nanoporous materials, for example in fluid mixture separation, gas capture, and heterogeneous catalysis, soft porous crystals can have specific applications as nanosensors and actuators, in energy storage and as molecular springs and shock absorbers. 	Moreover, they can be used --- a single crystals or as part of nanocomposite materials --- to create systems with counterintuitive or ``anomalous'' physical properties, like negative linear compressibility (in which the material under compression expands along certain directions, while undergoing a reduction in volume)\cite{Cairns2015} or negative adsorption (release of adsorbed molecules upon increasing gas pressure).\cite{krause2016}

In this Feature article, we highlight the recent progresses made in the computational characterization of soft nanoporous crystals and the prediction of their physical and chemical properties, focusing in particular --- but not exclusively --- on MOFs. This is not intended to be a thorough review of the modeling techniques that can be applied to these materials, but to illustrate the breakthroughs, evolution and perspectives in this field. For a background introduction, we refer the reader to the comprehensive reviews of \citeauthor{Odoh2015} and \citeauthor{Coudert2016}.\cite{Odoh2015, Coudert2016}.

\section{From local to global framework flexibility}

The first and simplest approach to molecular simulation of guest adsorption in nanoporous materials is to consider the host matrix to be rigid or frozen, acting only as an external field on adsorbate molecules. This has been considered the state-of-the-art approach for many years,\cite{Fuchs2001, Smit2008} and is employed in molecular simulations of phenomena such as: pure component adsorption and fluid mixture coadsorption, using grand canonical Monte Carlo (GCMC) simulations; transport and diffusion inside the pores using molecular dynamics (MD) methods; reactivity and catalysis using quantum chemistry calculations of representative clusters, where the rest of the structure is assumed to be \revadd{rigid}. This approach was relatively well justified when the most studied nanoporous materials were inorganic materials, such as zeolites: alumino-silicates with an immense diversity of shapes and topologies, and applications of high economic importance, e.g., in the oil industry. Due to their high stiffness --- quantified by a high Young's modulus, typically of the order of 100~GPa --- zeolites are hard to deform. For this reason, most of the initial studies of zeolites used a \emph{frozen} representation of the framework, specifically in GCMC and MD simulations. The impact of framework flexibility was observed in unusual cases of subtle structural transitions (like the adsorption of halocarbons in silicalite-1\cite{Jeffroy2008}) or for the diffusion of molecules through very narrow pores and windows.\cite{Leroy2004}

\subsection{The need to go beyond the frozen framework picture}

\begin{figure}[htbp]
	\centering
	\includegraphics[width=8cm]{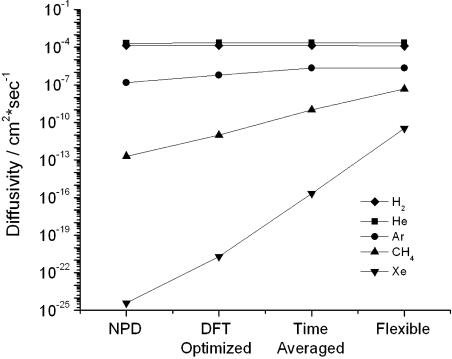}
    \caption{Diffusivities in the infinite dilution limit for five gases (hydrogen, helium, argon, methane and xenon) with four different computational models of the ZIF-8 framework at 300~K. Figure taken from Ref.~\citenum{haldoupis2012}.}
    \label{fig:haldoupis2012}
\end{figure}

However, the past decades have seen a rise in novel classes of nanoporous materials, including MOFs, COFs, SOFs, and other weakly-bound molecular crystals. Based on organic linkers, metallic center, coordination bonds, hydrogen bonds and $\pi$--$\pi$ stacking interactions these material are softer and easier to deform than zeolites.\cite{Coudert2015} The rigid representation of the framework quickly shows limited accuracy when applied to these flexible nanoporous materials. One of the first studies highlighting this limitation in detail was a study by \citeauthor{haldoupis2012}\cite{haldoupis2012} on the diffusion of small molecules a zeolitic imidazolate framework (ZIF). ZIF-8 has a zeolite-like structure, where the silicon atoms are replaced with a metal and the oxygen atoms with an imidazolate linker. The ZIF-8 framework specifically presents a sodalite topology, with pores connected by 6- and 4-member ring windows. In their work, \citeauthor{haldoupis2012} realized that molecules with a larger diameter than the pore window could diffuse through the structure. This led to the study of flexibility on the diffusion coefficient of small molecules. Their results are illustrated in Figure~\ref{fig:haldoupis2012}, showing that a fully flexible description of the framework is needed to reproduce the experimental diffusion data. The same effect is observed in adsorption, as ZIF-8 can adsorb butane, which would not fit within its pores by purely conventional geometric criteria based upon a rigid framework structure.

\subsection{\revadd{Describing flexibility: \abinitio or classical forcefields}}

To account for flexibility in molecular simulations, one relies on a description of the potential energy of the nanoporous framework as it deforms and the atoms move. The primary methods of describing this energy are \abinitio (or ``first principles'') computations, i.e. solving the Schr{\"o}dinger equation for each conformation of the framework; or classical forcefields, a set of parametrized functions describing the framework energy as a sum of specific contributions: bonds stretching, angles bending, etc. Quantum chemistry calculations can be more precise, but their computational cost and scaling for large number of atoms is very expensive. Classical forcefields are cheaper, but suffer from two issues: first the energy they predict is less precise, however, many still reproduce thermodynamic behavior. Second, they require parametrization for the specific system of interest. Transferable forcefields can be transferred from one system to another, i.e. used on systems other than the one they were developed for, but with a reduced accuracy.

When choosing a classical forcefield for a flexible structure, two strategies are currently used. One can take advantage of the fact that MOFs and related materials are constructed from organic and inorganic compounds, which there exists many nice and transferable forcefields, such as the general AMBER forcefield (GAFF),\cite{Wang2004} or the Universal forcefield (UFF).\cite{Rappe1992} To improve accuracy these forcefields can be modified for a specific material by setting the charges and bending values around the metal center. For example, \citeauthor{zheng2012} \textit{et al.} used values from AMBER forcefield and charges around the metal center from DFT calculations to describe ZIF-8 flexibility\cite{zheng2012} and Heine \textit{et. al} have tuned UFF for use in MOF structures.\cite{Addicoat2014} Alternatively, machine learning techniques can be employed to create a non-transferable but more precise classical forcefield from \abinitio DFT computations. This is a somewhat new development in the field of computational chemistry, and it is for example the approach taken by MOF-FF\cite{sareeya2013} and Quick-FF\cite{vanduyfhuys2015}. These forcefield generators use results from DFT and machine learning to derive a new classical forcefield for the specific system of interest. These forcefields can include a large number of analytical terms with complex functional forms and numerous parameters, as the fitting of these parameters is not done explicitly but instead handled through a multivariate optimization algorithm.

\revadd{When a classical forcefield does not exist, or when the existing parameterization fails to reproduce important physical properties of the system, \abinitio calculations can be used to compute the energy and forces acting on the system --- typically in the Density Functional Theory (DFT) approach. Such \abinitio simulations make fewer assumptions on the nature of the system, and do not use \emph{ad hoc} empirical parameters as the classical forcefields do. Moreover, they provide a description of the electronic degrees of freedom of the system, and thus allow one to calculate electronic properties in the ground or excited states. However, this accuracy and transferability come with a high computational cost and with a few other drawbacks. Because the computation of the energy and interatomic forces is more expensive, simulation is restricted to rather small systems (hundreds of atoms) and to typically shorter simulation timescales: from 10 to 100 picosecond trajectories for molecular dynamic, compared to 10--100 nanoseconds for classical simulations. This means that the simulation might not be completely converged toward equilibrium, and that results can carry a bigger statistical uncertainty. Secondly, while the calculated energies and forces can be more accurate with \abinitio methods, the choice of methodology and parameters (exchange--correlation functional, dispersion corrections, ...) is crucial, especially for nanoporous materials whose behavior is driven by a balance of relatively weak interactions. As an example, Haigis \emph{et al.} have shown that in MIL-53(Ga),\cite{haigis2014} the dispersion correction method used has a strong influence on the convergence and result of constant-pressure simulations. Nevertheless, \abinitio methods are attractive for molecular simulation of soft porous crystal because they are applicable to a wider range of structures than classical methods --- which need to be parametrized for every new structure. A typical example of the use of \emph{ab initio} molecular dynamics in soft porous crystals is the investigation by Chen \emph{et al.}\cite{chen2013} of the structural transitions in MIL-53(Sc) upon carbon dioxide adsorption and temperature changes, allowing the determination of the structure of novel phases of the material and fundamental insights from the molecular-level interactions further into the origin of the breathing transition.}

\bigskip
\section{Phase transitions and large-scale structural changes upon pressure or adsorption}

Flexible nanoporous materials are able to deform under the application of stress or mechanical load, which can come from either the outside of the structure (e.g. by nanoindentation or compression by a pressure-transmitting fluid) or from the guest molecules (adsorption-induced stress).\cite{Neimark2011} Indeed, gas adsorption in these flexible structures can have the same effect on the structure as an external stress.\cite{gor2017} At low loading adsorbed gas interactions with framework walls, equivalent to a positive external stress, causes the structure to contract. Subsequently, at higher loading gas movement and collisions with the framework, equivalent to negative external stress, usually cause the material to expand.

Deformations can remain in the elastic regime or induce large scale nonlinear transformations of the structure depending on the magnitude, the nature of the stress and the mechanical properties of the framework structure. A common example of such transformations is the occurrence of pressure-induced or adsorption-induced phase transitions. For example, the MIL-53 family of MOFs\cite{Serre2002} exhibit a \emph{breathing} phenomenon which is caused by a combination of two factors: two or more phases exist with different porosity, and the transitions between them is driven by adsorption-induced stress. For the MIL-53(Al), the two structures are an open pore structure with high pore volume and a closed pore structure with lower pore volume. Initially, the empty structure is more stable in the open pores phases however, at low guest loading the closed pore phase is more stable. Increased loading will increase the stability of the open-pores phase again and during adsorption the structure will transform from open to closed and back to an open phase.

\subsection{Molecular dynamics simulation}

Molecular simulation of adsorption is typically conducted in the grand canonical ensemble, using GCMC. In this ensemble, the chemical potential of the adsorbate is fixed, and specific Monte Carlo moves are used to insert or remove particles in the system. Metropolis Monte Carlo is a crucial method for this application as it allows the simulation of systems with a varying number of particles, as long as the Metropolis criterion is well defined.

This causes a problem for simulating large structural changes in a system induced by adsorption. For efficiency reasons, Monte Carlo simulations move few particles in each trial move. Moving many particles in a random way could lead to frequent rejections and increase the simulation time required to equilibrate the system. However large scale changes in framework structure are caused by the collective displacement of framework atoms. The challenge of simulating the coupling between adsorption and framework deformation has spurred methodology development.

Another standard molecular simulation method is MD, which integrates the movement of the particles in the phase space following Newton's equation of motion. Thus MD is excellent at simulating collective behavior and many new methods employ MD in order to reproduce collective behaviors.

One of the first approaches used by the community to simulate this coupling is to introduce grand canonical features during an MD simulation. In grand canonical molecular dynamics (GCMD),\cite{eslami2007} a particle is flagged as ``partial'' and slowly inserted or removed from the system by scaling its mass and interaction with the system. Another possible simulation scheme is to insert or delete molecules periodically from the system using a Metropolis criterion. After a fixed MD simulation time $\Delta t_{MD}$, molecules are randomly inserted or removed.\cite{stockelmann1999} A third proposition in the literature is to use mixed simulations; \emph{i.e.} run a short MD simulation in the $(N, P, T)$ ensemble, then a short MC simulation in the $(\mu, V, T)$ ensemble and continue until convergence.\cite{zhang2013}

\begin{figure}[ht]
	\centering
	\includegraphics[width=0.8\linewidth]{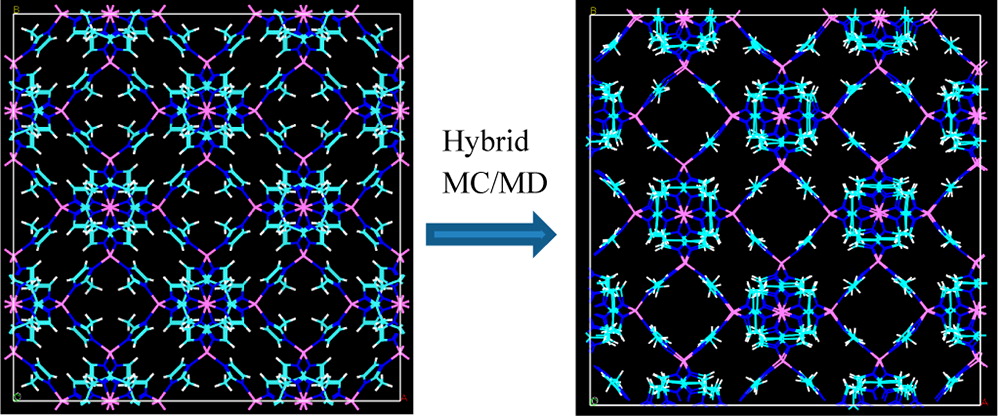}
    \vskip 1ex
	\includegraphics[width=0.5\linewidth]{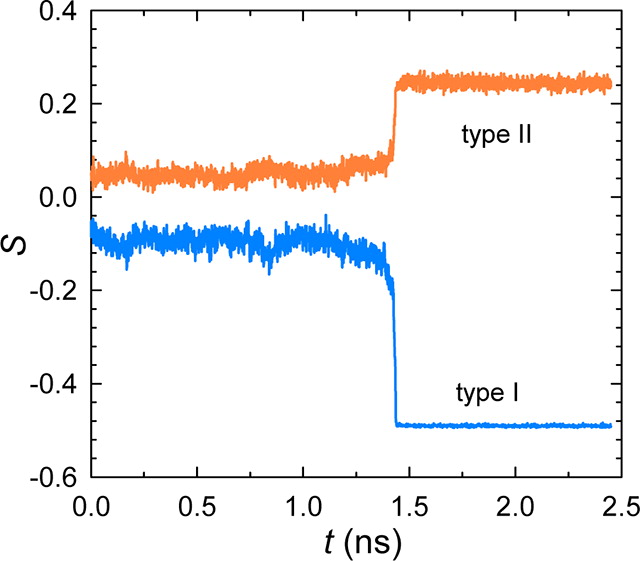}
    \caption{Study of the continuous deformation of ZIF-8 during nitrogen adsorption with mixed MD/MC simulation. Figure reproduced from Ref.~\citenum{zhang2013} Top is the structure of the framework at low (LP phase) and high (HP phase) loading, bottom is the transition from LP to HP as observed during a MD run.}
    \label{fig:zhang2013}
\end{figure}

For example, the ZIF-8 framework goes from a low pressure (LP) structure to a high pressure (HP) structure under loading. The difference between the two structures is a rotation of the imidazolate linkers. \citeauthor{zhang2013} describe the opening of ZIF-8 structure by adsorption using a mixed MD/MC simulation as illustrated in Figure~\ref{fig:zhang2013}.\cite{zhang2013} This study found a mixed MD/MC simulation scheme, in conjunction with specific forcefields for ZIF-8 and N$_2$ , able to reproduce the unique isotherm.

An issue with methods employing MD simulations is that it is unknown whether the ergodic hypothesis hold for simulations with a varying number of particles. Ergodicity is the property in which the time-averaged value of any physical quantity of the simulated system is equal to the thermodynamic average of this property over the accessible phase space; which is the real-world value of the average. The question whether MD simulation with a non-constant number of particles can actually generate the right ensemble average and whether the ergodic hypothesis holds in this case is an open research question.\cite{dellesite2016} Notably, as Monte Carlo simulations directly sample the phase space, they do not suffer from this problem.

\subsection{Monte Carlo simulation}

To be certain that we are sampling the right ensemble an alternative approach uses only Monte Carlo methods and there are a number of methods that have been employed to simulate framework flexibility with reasonable efficiency.

\begin{figure}[ht]
	\centering
	\includegraphics[width=0.8\linewidth]{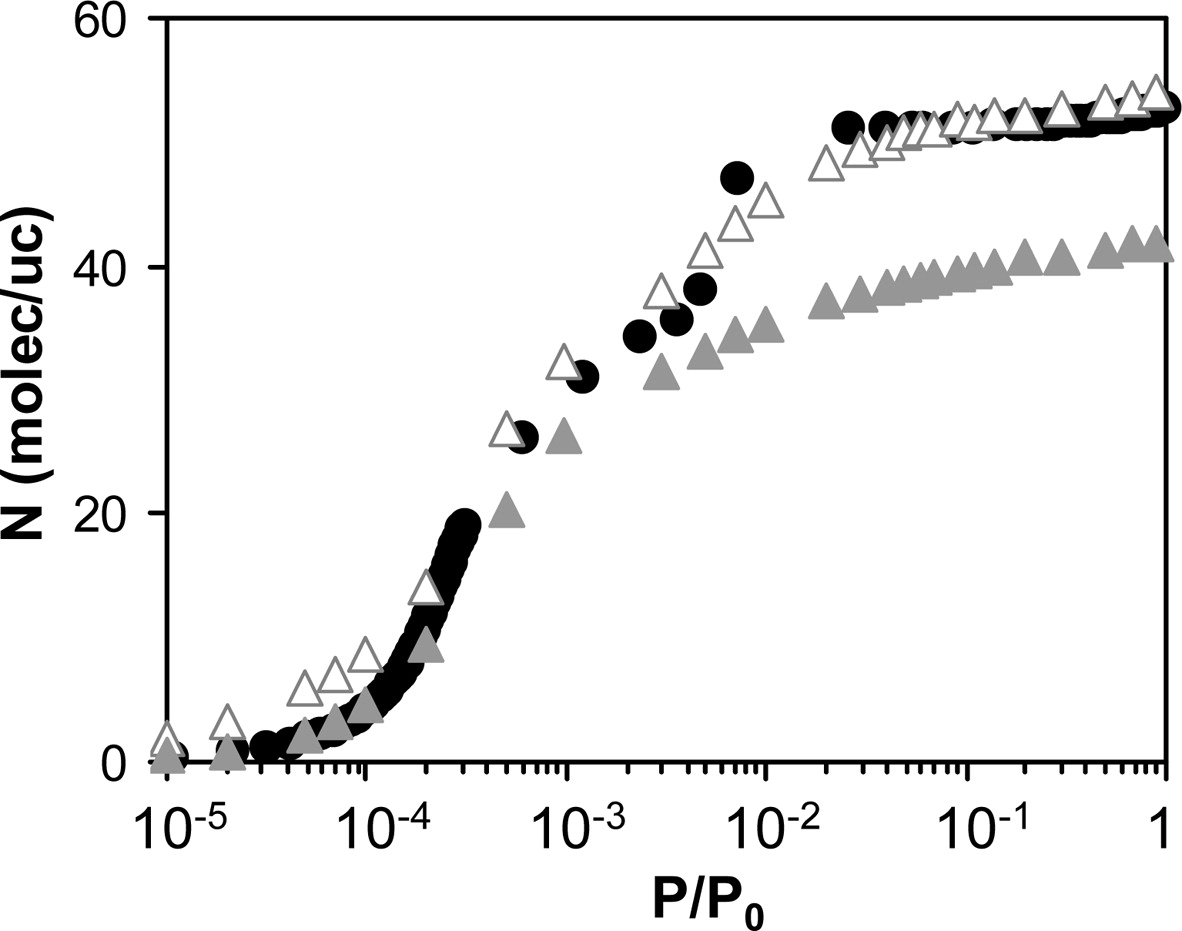}
    \caption{Experimental (circles), GCMC in LP phase (gray triangles) and GCMC in HP phase (open triangles) isotherms for adsorption of N$_2$ in the ZIF-8 framework. Figure reproduced from Ref.~\citenum{fairen-jimenez2011}.}
    \label{fig:fairen-jimenez2011}
\end{figure}

For systems with multiple stable states but no flexibility between the states (for example systems with a first order phase transition) it is possible to perform several GCMC simulation in the rigid states and then reconstruct the whole simulation by mixing results from different phases. For example, \citeauthor{fairen-jimenez2011} also studied the adsorption in ZIF-8,\cite{fairen-jimenez2011} were able to reproduce the isotherm at low loading using a simulation in the rigid LP framework and subsequently simulate the high-loading portion using the rigid HP structure as displayed in Figure~\ref{fig:fairen-jimenez2011}.

To efficiently simulate framework flexibility the collective behavior of the framework atoms must be reproduced. One possible approach uses a short MD trajectory to generate a new trial configuration; and then accept or reject this new configuration using the usual rules for Metropolis Monte Carlo. This MD trajectory will sample collective movements and, because the MD integrator preserve the energy of the system, most of the trial moves will be accepted. This is the hybrid Monte Carlo (HMC) scheme, first proposed in \citeyear{duane1987}\cite{duane1987} and applied to material science in \citeyear{horowitz1991}.\cite{horowitz1991} In this scheme, the MD integrator must include specific properties to ensure the generated Markov chain is ergodic and micro-reversible. Specifically, the integrator needs to preserve the volume of the phase space (be symplectic) and to be time-reversible. A $(N, V, T)$ MD simulation can be used to sample flexibility of the structure when there is no volume change involved. However, when volume of the framework changes due to adsorption a $(N, P, T)$ simulation is required to account for these volume changes. Unfortunately, most standard $(N, P, T)$ integration algorithms are not volume preserving and very few are time-reversible. Currently, there is a research effort to generate an isotherm-isobar algorithm with all these properties.\cite{ciccotti2001, yu2010}

We want to \revadd{emphasize} here that HMC and mixed MD/MC simulations are fundamentally different. In HMC, the MD simulation is used to generate a new state in the Monte Carlo Markov chain and the whole simulation follows the usual Metropolis Monte Carlo scheme. While in mixed MD/MC simulations, the method used to generate a new configuration is switched from MD to MC without any theoretical foundation.

\begin{figure}[ht]
	\centering
	\includegraphics[width=\linewidth]{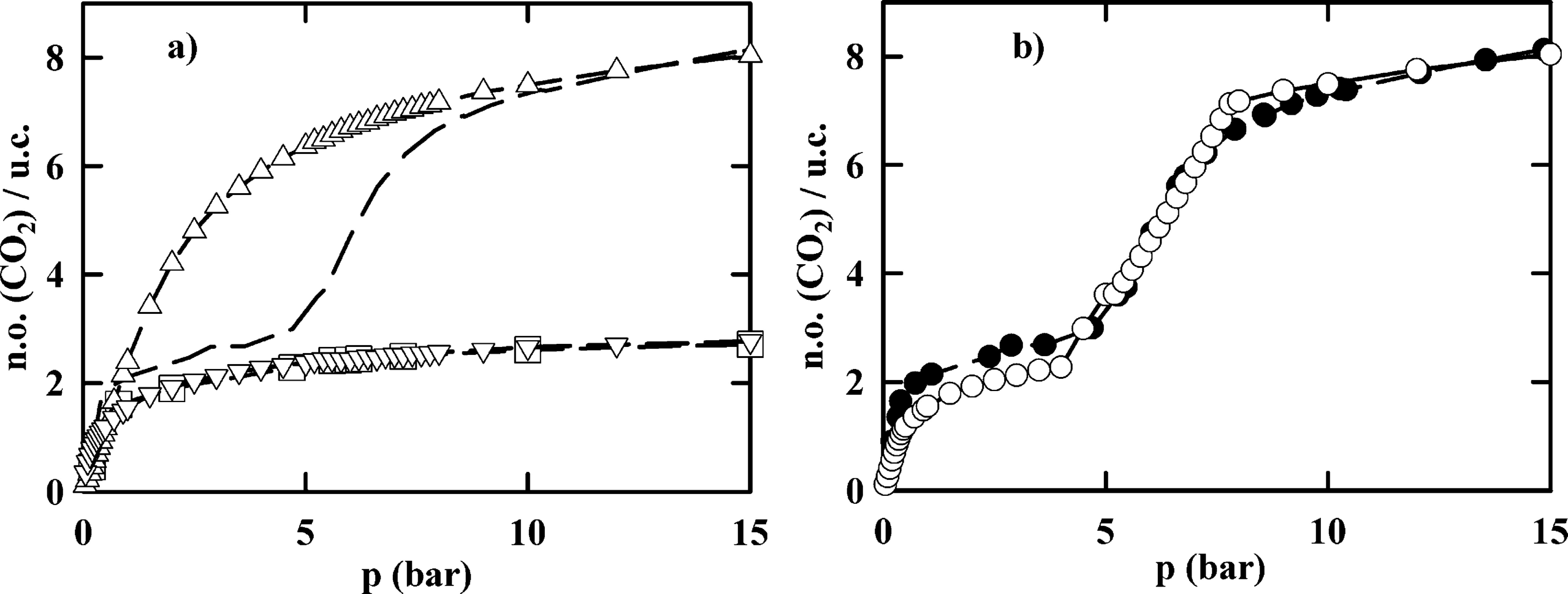}
    \caption{Adsorption isotherms of CO$_2$ in MIL-53(Cr) at 300 K: (a) experimental data ($\medblackcircle$), hybrid GCMC simulations ($\Box$), rigid Lp ($\medtriangleup$), and Np ($\medtriangledown$); (b) experimental data ($\medblackcircle$) and the combination of H-GCMC and ``phase mixture'' model ($\medcircle$). Figure reproduced from Ref.~\citenum{ghoufi2010}.}
    \label{fig:ghoufi2010}
\end{figure}

HMC has been used in combination with a mixing parameter to reproduce  breathing of the MIL-53(Cr) by \citeauthor{ghoufi2010}.\cite{ghoufi2010} This study observed that a hybrid GCMC simulation effectively reproduced the first transition but failed to reproduce the re-opening of the structure. However using a phase mixture of the two phases of transition the simulation reproduced the second transition and the full isotherm as illustrated in Figure~\ref{fig:ghoufi2010}.\footnote{We note in passing that \citeauthor{ghoufi2010} used the Berendsen (also called weak coupling) barostating algorithm for the HMC step, although the reference used to justify that this algorithm is volume preserving and time reversible\cite{morishita2000} does not discuss these properties: Ref.~\citenum{morishita2000} does not even discuss the barostating version of the weak coupling algorithm, but focuses only on the thermostat. This paper demonstrates that the Berendsen thermostat does not sample either of the canonical or micro-canonical ensemble, but rather something between; and that the coupling parameter allows switch between these two ensembles continuously.}

\subsection{Free energy methods}

Molecular simulation methods are useful to obtain an atomistic representation of the system and the chemical process occurring. However, it is speculative to interpret the results of a single simulation as a \emph{movie} of the simulated process. The events of a simulation is only one realization of the possible trajectories. Results obtained by MD rely on the fact that all the realizations are statistically significant, and thus the average along the trajectory is a statistical average: the ergodic hypothesis.

\begin{figure}[ht]
	\centering
	\includegraphics[width=0.9\linewidth]{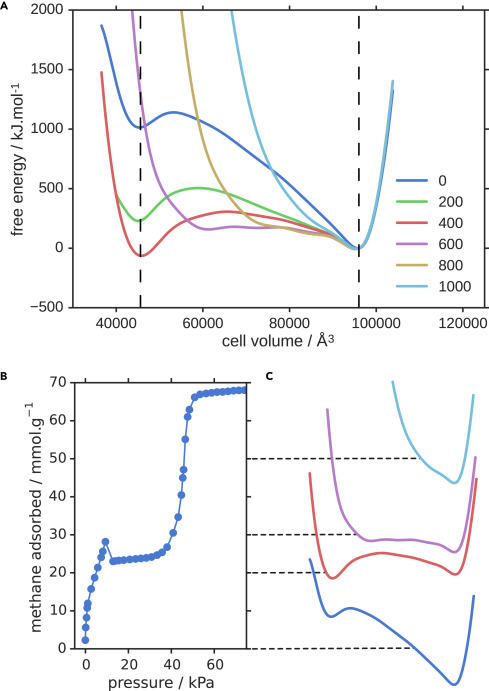}
    \caption{(A) Free-energy profiles obtained from $(N, V, T)$ simulations as a function of unit-cell volume at 120~K with increasing amounts of methane molecules (molecules per unit cell). Volumes corresponding to the open and close pores phases are displayed as dashed lines \revadd{and plain lines correspond to free energy at different loadings (indicated in molecules per unit cell)}. (B and C) The methane isotherm for DUT-49 at 111 K (B) is shown with the corresponding free energy profiles (C) \revadd{for selected values of the methane loading, taken from panel (A).} Figure reproduced from Ref.~\citenum{evans2016}.}
    \label{fig:evans2016}
\end{figure}

Obtaining an overall view from MD simulation requires understanding at the thermodynamic level using free energy methods.\cite{Rogge2015} These methods range from simple thermodynamic integration to more complex meta-dynamic methods.\cite{Frenkel2002} A recent study of DUT-49 highlight these methods to study the flexibility of porous materials. This MOF has been shown to present a \emph{negative gas adsorption},\cite{krause2016} \emph{i.e.} there is a part in the adsorption isotherm with negative slope, where increasing gas pressure result in desorption of gas illustrated in Figure~\ref{fig:evans2016}. \citeauthor{evans2016}\cite{evans2016} used thermodynamic integration of multiple simulations at constant volume and increasing guest loading of methane to explain this behavior. The free energy profiles obtained are reproduced in Figure~\ref{fig:evans2016}. The structure transforms from a bi-stable system with a minimum at high cell volume at 0 loading, to a bi-stable system with the minimum at low cell volume at 400 molecules loading and finally to a mono-stable system at high cell volume for 800 molecules. In this report negative gas adsorption was attributed to an abrupt transition between the stable high volume phase and the meta-stable low-volume phase.

\subsection{Thermodynamic studies}

In addition to molecular simulation, thermodynamic equations can be solved either analytically or numerically to describe the flexibility of these systems.

The thermodynamic ensemble of choice for studying coupled adsorption and deformation is the Osmotic ensemble,\cite{escobedo1998} where the thermodynamic potential $\Omega$ is defined at fixed number of atoms in the host $N^h$, temperature $T$, pressure $P$ and chemical potential of the guest $\mu^g$ by:
\[\Omega(N^h, T, P, \mu^g) = F^h(N^h, T, P) + P V^h - \int_0^P V_m^g n^g(\mu^g, p, T) \text{d} p \ ;\]
where $F^h$ and $V^h$ are the host free energy and volume respectively, $V_m^g$ is the guest molar volume and $n^g$ the quantity of matter of the guest. For multiple guests, the same definitions apply with a sum over the guests in the integral.

This expression is useful for co-adsorption prediction in flexible structures. Co-adsorption is the study of simultaneous adsorption of multiple components in a single structure and is at the basis of gas separation studies and application. Considering a structure presenting phase transitions, but under the hypothesis that the various phases are rigid, then we can compute the transition pressure and temperature for a given composition of the adsorbing mixture knowing only the single-component isotherms. One method which employs this is the Osmotic framework adsorbed solution theory (OFAST) method.\cite{coudert2010} In this method, knowledge of the single component isotherms is used to extrapolate isotherms in all the structure phases; subsequently the potential $\Omega$ is minimized to find the more stable phase at a given composition, pressure, temperature. Once the most stable phase is known standard methods such as ideal adsorbed solution theory (IAST), GCMC, or any method which employed a rigid framework can be used to predict the co-adsorption data.

\begin{figure}[ht]
	\centering
	\includegraphics[width=0.9\linewidth]{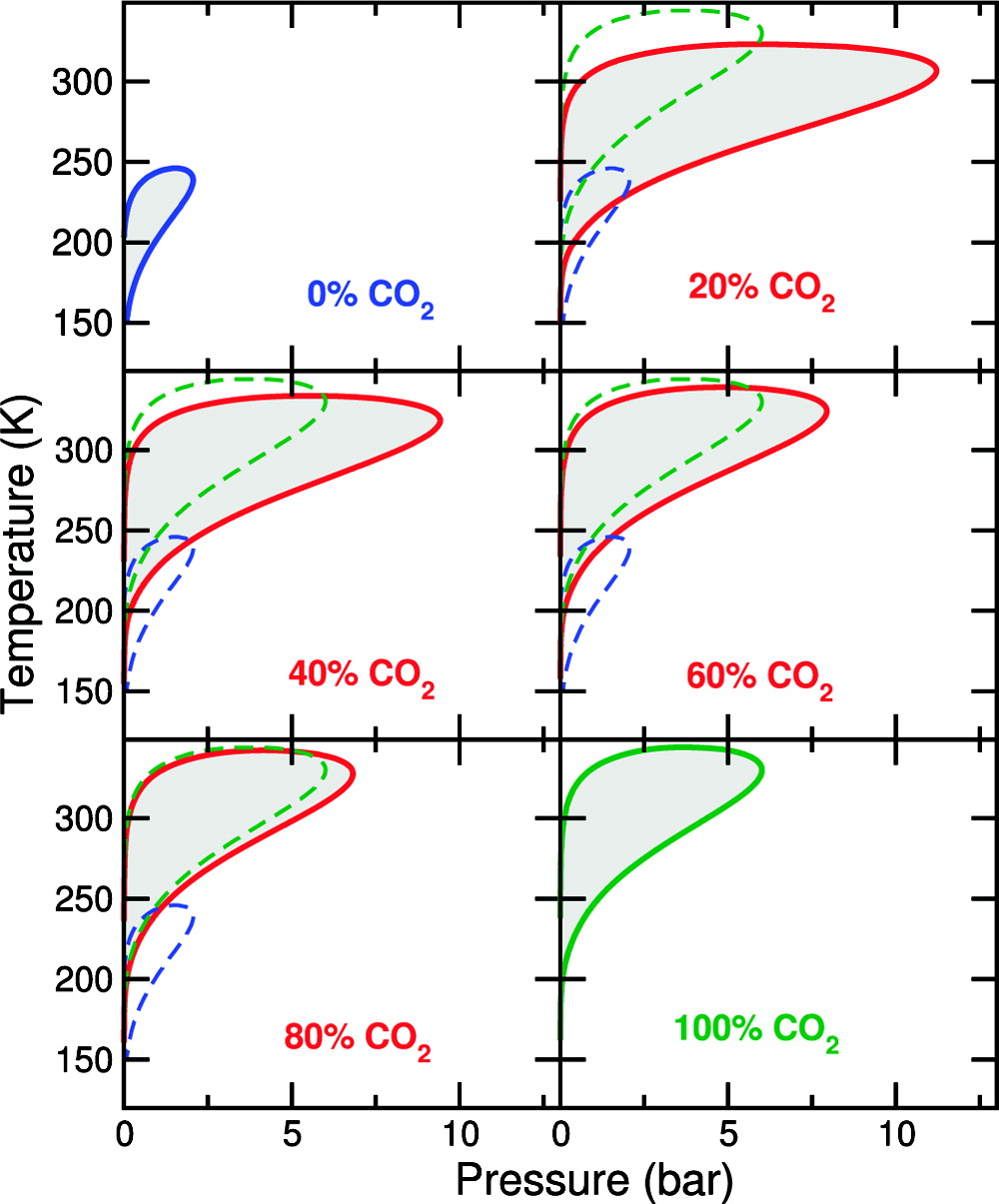}
    \caption{Temperature--pressure phase diagram of MIL-53(Al) upon adsorption of a CO$_2$/CH$_4$ mixture, with increasing CO$_2$ molar fraction. Dashed lines correspond to pure component diagrams. Figure reproduced from Ref.~\citenum{ortiz2012}.}
    \label{fig:ortiz2012}
\end{figure}

The OFAST method has been used to study the phase diagram of the breathing phenomenon of MIL-53(Al), from only the experimental single-component isotherms.\cite{ortiz2012} \citeauthor{ortiz2012} demonstrated that for some compositions of the gas mixture (around 20\% CO$_2$/ 80\% CH$_4$), the closed pore phase had a much greater stability range in the temperature-pressure diagram compared to pure phase adsorption illustrated in Figure~\ref{fig:ortiz2012}).

However, this osmotic potential can be difficult to determine as it relies on the free energy of the host structure. To study a fully flexible structure the variation of the free energy with the structure deformation must be known. Recently, \citeauthor{zang2011} studied the structural transition in alumino-silicate nanotubes (called imogolites) upon adsorption.\cite{zang2011} The velocity density of states with a quasi-harmonic approximation was used to compute the free energy of a system, and then OFAST was applied using this free energy expression.

\subsection{Description at larger scale}

\revadd{We mention in this section a few alternative approaches for the treatment of flexibility in nanoporous crystals, which differ from the more conventional approaches of direct molecular simulation, by taking a somewhat larger scale, using either mesoscopic or macroscopic description of the systems. One such approach is that of \emph{poromechanics}, which relies on the equations of continuum mechanics to study the behavior of fluid-saturated porous media.\cite{coussy2005} In this approach, based on macroscopic laws, the solid--fluid interactions and confinement effects are described implicitly by an interaction energy and ``interaction pressure'',\cite{PijaudierCabot2011} which can be derived from classical adsorption models such as the Langmuir isotherm. Calculations based on poromechanics (as described in some detail in the recent review by Gor et al.\cite{gor2017}) have been shown, for example, to provide a good description of the deformation of coal induced by CO$_2$ and CH$_4$ adsorption\cite{Vandamme2010} as well as the modifications in the stiffness of the materials upon fluid uptake.}

\revadd{Finally, we mention here the emergence of statistical models for the description of flexible nanoporous materials at scales larger than a few unit cells. This is exemplified in the recent work by Simon et al,\cite{Simon2017} where the authors developed a statistical mechanical model of gas adsorption in a model porous material where a rotating ligand moiety is shared between porous cages. This model captures a rich phenomenology of the material upon adsorption, where the interplay between host dynamics and guest adsorption can lead to inflections, steps, and hysteresis in the adsorption--desorption isotherms. Along a similar line, a simple Hamiltonian-based model was proposed by Triguero et al.\cite{Triguero2012} for the ``breathing'' structural transition of MIL-53 at the scale of a whole crystal, based on elastic compatibility equations, classical adsorption laws and elastic cell--cell interactions.}

\bigskip
\section{Defects, disorder, and framework breakdown}

In addition to local flexibility of their framework and stimuli-induced large-scale structural transitions, soft nanoporous crystals are also prone to feature defects in their structure, as well as long-range disorder.\cite{Bennett2017} Moreover, it is more and more being recognized that defects and disorder --- like flexibility --- do not necessarily impact negatively the physical and chemical properties of a porous material, but can also introduce novel desirable behavior and improve performance for some functions: improving adsorption affinity or uptake, augmenting catalytic activity, etc.\cite{Sholl2015, Cheetham2016} The same is true for amorphous porous materials, which may present specific functionality while retaining characteristics of their crystalline counterparts: for example, improved mechanical stability while remaining porous.\cite{Bennett2014} In this section, we highlight some of the recent computational advances in addressing the presence of defects and disorder in nanoporous materials.

\subsection{Characterization of defects and their formation}

First and foremost in the use of computational methods to understand defects and disorder in metal--organic frameworks is the characterization of their structure, their formation mechanism, as well as one of their most sought-after property: their catalytic activity. Along with X-ray diffraction techniques -- often used \emph{in situ} -- molecular simulation offers detailed insight into framework defects at the molecular level. It has thus been widely used, for example, to understand the nature of missing linker defects in materials of the UiO-66 family, probably the most studied system when it comes to defects in MOFs. Indeed, UiO-66 has a relatively high concentration of missing linker defects, which can furthermore be finely tuned by the presence of modulator during the synthesis of the material. The presence of these defects has been correlated with catalytic activity, making them beneficial for potential applications in this area. However, despite this fact having been highlighted in detail since 2013,\cite{Wu2013} there is still no consensus on the local structure of those defects.\cite{Trickett2015, Ling2016}

We will highlight here two recent computational advances in our understanding of defect formation and structure in porous metal--organic frameworks. The first one is a study, by Zhang et al.,\cite{Zhang2016} examining the structure and stability of several point-defect structures in the zeolitic imidazolate framework ZIF-8, a nanoporous Zn(2-methylimidazole)$_2$ framework with sodalite (\emph{sod}) topology. By using static quantum calculations at the density functional theory (DFT) level, the authors calculated the local structure and formation energies for various hypothetical point defects (including vacancies, substitutions, and dangling linkers). They showed that several of the defect structures considered have relatively low energy difference with the defect-free crystal, and also characterize the energy barriers to the defect formation process.

In another work, studying missing linker defects in UiO-66, Ling et al.\cite{Ling2016} have combined static DFT calculations and first principles molecular dynamics in order to propose a detailed characterization of the local structure of the defects. They characterize the position and coordination mode of charge balancing hydroxide anions, and demonstrate a strong dynamic behavior at the defect site, namely a rapid proton transfer between the hydroxide anion and extra-framework physisorbed water molecules. From the methodological point of view, this is an interesting development as it clearly shows the insight gained from molecular dynamics simulations --- here at the quantum chemical level --- which could not be obtained from purely static calculations, although these are more often used in studies. From the practical point of view, it also has important consequences on the nature of the defects in the UiO-66 material, as the defect centers show increased acidity and can act as a source of highly mobile protons, conferring the MOF an increased potential for catalytic functionality.

\subsection{Impact on catalytic activity and properties}

Indeed, in addition to the characterization of their structure and formation energies, a large part of the computational literature on defects in MOFs has dealt with their impact on catalytic properties of the materials. The Van Speybroeck group, in particular, has published several studies of mechanistic pathways for catalytic reactions on UiO-66 structures with defects, by means of nudged elastic band (NEB) calculations,\cite{Vandichel2016} on reactions such as citronellal cyclization\cite{Vandichel2015} and Oppenauer oxidation of primary alcohols\cite{Hajek2017}. Missing linker defects, as well as linker modification and node metal substitution, have also been shown to provide a pathway to tune MOFs such as UiO-66 for photo-catalytic purposes.\cite{DeVos2017}

\begin{figure}[t!]
	\centering
	\includegraphics[width=0.8\linewidth]{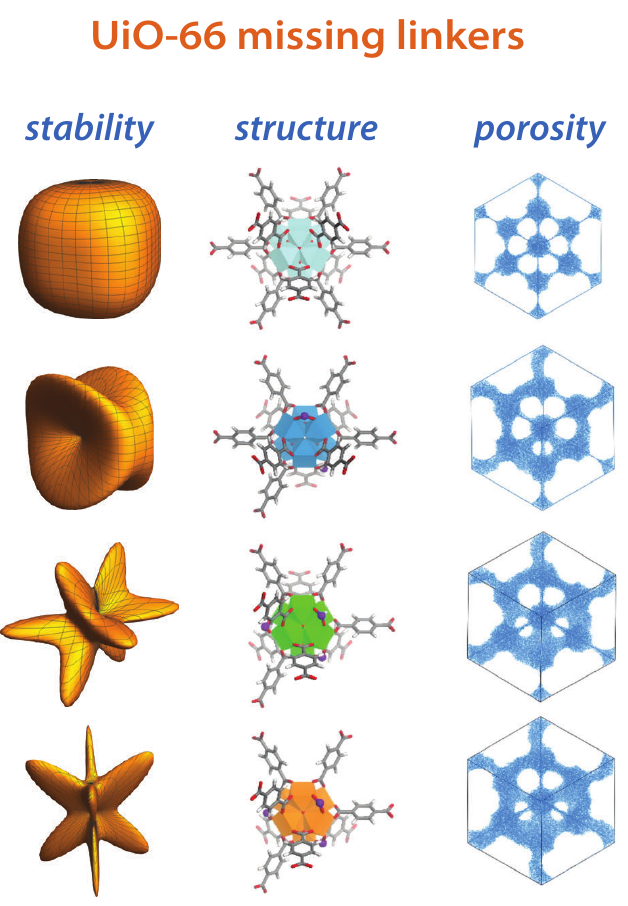}
    \caption{Impact of the presence of defects in the UiO-66(Zr) family of MOFs: the anisotropy of directional Young's modulus as a 3D surface (left column) and accessible microporosity (right column) both depend on the local structure (middle) and  the number of missing-linker defects, which increases from top to bottom. Figure adapted from Ref.~\citenum{Thornton2016}.}
    \label{fig:defects}
\end{figure}

The impact of the presence of defects on physical properties of MOFs has also been studied, especially in the case of adsorption. Defects typically tend to increase adsorption capacity and affinity,\cite{Wu2013, Shearer2016} by providing larger pores, more internal surface, introducing polar or hydrophilic groups, lowering the symmetry and thus increasing the electric field, and providing access to undercoordinated metal sites. But defects impact other physical properties of MOFs, and there has recently been some focus on properties such as thermal expansion\cite{Cliffe2015} and mechanical stability.\cite{Thornton2016} In the case of missing linker defects in UiO-66(Zr), for example, Thornton et al.\cite{Thornton2016} showed that defects increase carbon dioxide uptake at high pressure, where the formation of larger cavities lead to higher pore volumes and CO$_2$ uptake than the perfect defect-free crystal structure. However, mechanical stability is compromised upon increase in defect concentration, as increased porosity (reduced density) leads to lower elastic moduli --- and in particular lower shear moduli.

\subsection{Defects and disorder}

In metal--organic frameworks, as in other ``conventional'' solids, the presence of defects is intrinsically linked with the potential for disorder in the structure.\cite{Bennett2017} However, the question of disorder in MOFs is not widely addressed in the published literature. On the experimental side, the main tool for MOF characterization is the crystallographic structure obtained, e.g., from experimental X-ray diffraction data. On the theoretical side, both structural studies and adsorption studies are usually performed on ideal structures where possible disorder has been removed, by tools that do not allow to take into account dynamic disorder (such as energy minimization or rigid structure calculations). Yet, for macroscopic properties averaged overall an entire crystal (such as adsorption isotherms or unit cell parameters), these methods provide a reasonable description. However, there are some cases where the presence (and extent) of disorder within the solid frameworks need to be addressed.

One such example is the correlated nanoscale disorder of missing-linker defects in UiO-66 materials, which was experimentally shown to exist: these defects are not included in the framework in a random manner, but the position of one vacancy affects the likelihood of vacancy inclusion at neighboring sites, forming well-defined nanoscale domains in the crystal. A computational study at the density functional theory level later showed that this leads to an increased negative thermal expansion in UiO-66(Hf),\cite{Cliffe2015} allowing the thermomechanical properties of the MOF to be systematically tuned via controlled incorporation of defects. A second example is the recent study of the distribution of cations within the structure of bimetallic MOFs of the UiO-66 and MOF-5 families.\cite{Trousselet2016} Using a methodology based on a systematic study of possible cation distributions at all cation ratios by means of quantum chemistry calculations, Trousselet et al. showed that bimetallicity is overall more favorable for pairs of cations with sizes very close to each other, owing to a charge transfer mechanism inside secondary building units. On the other hand, for cation pairs with significant mutual size difference, metal mixing is globally less favorable due to unfavorable mixing-induced strains. This is an important new development in the computational field, given the growing number of polymetallic MOFs (also called \emph{multivariate} or \emph{heterogeneous} or \emph{mixed-metal} MOFs) reported and their potential for applications,\cite{Wang2014, Zou2016} as there is currently little experimental information on the ordering (or disordering) of cations in these complex systems.

\subsection{Reactivity of the framework and hydrothermal stability}

Finally, there are cases where disorder in the framework becomes so extreme that the crystalline nature of the MOF is lost. This can happened for example upon mechanical constraint or heating, leading to formation of an amorphous state. It can also be due to lack of stability of the framework upon reaction with a guest molecule, typically water. However, computational studies of these phenomenon --- and the modeling of complex amorphous structures in general --- is very expensive in terms of computational power, and there are thus relatively few studies on the topic. One possible exception of the refinement of structural models of amorphous frameworks, which can be done by Reverse Monte Carlo modeling of experimental X-ray (or neutron) diffraction data.\cite{Bennett2010}

\begin{figure}[t!]
	\centering
	\includegraphics[width=0.95\linewidth]{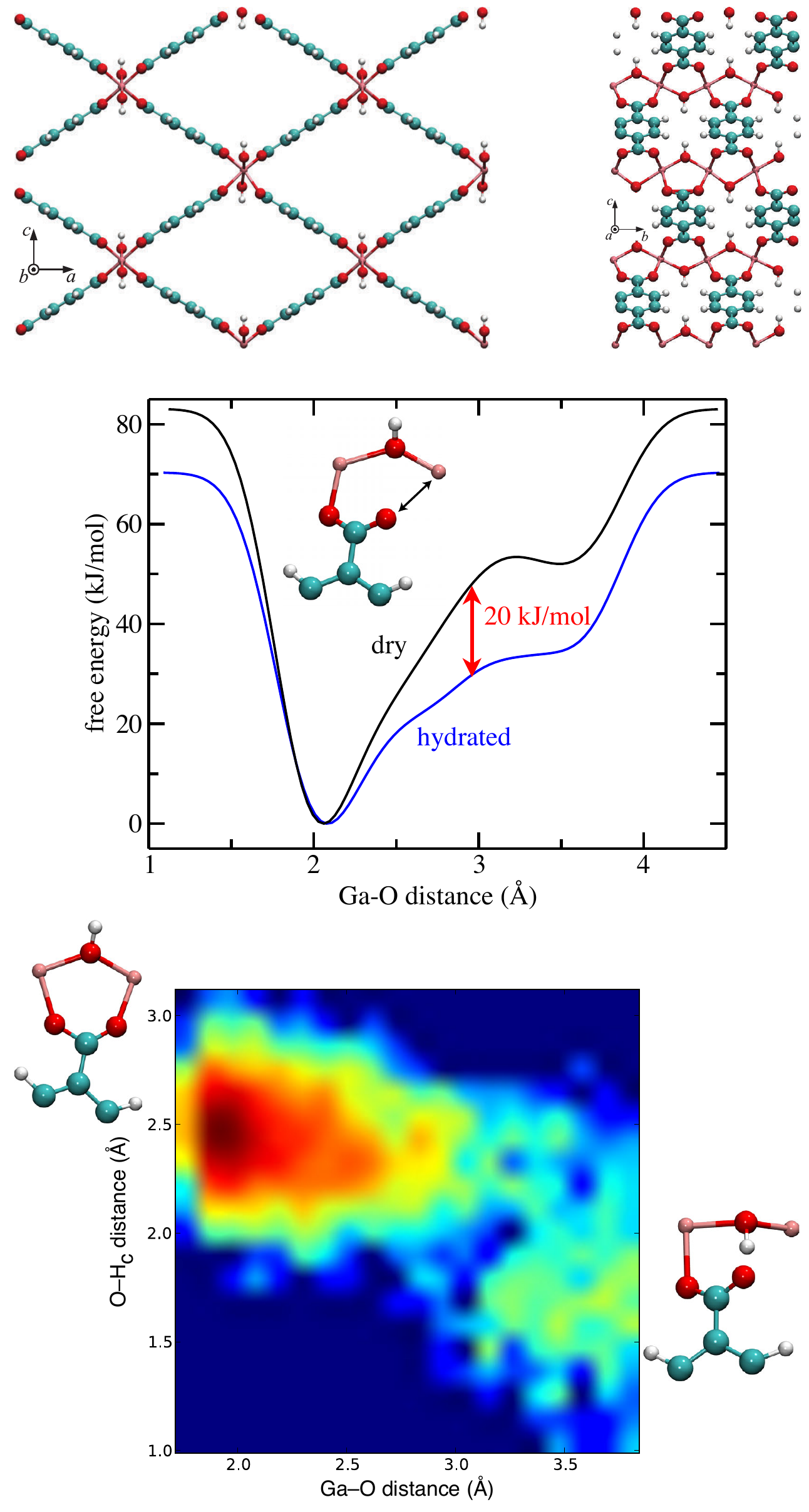}
    \caption{Top: views of the MIL-53(Ga) framework. Middle: Free energy profiles of Ga--O bond breaking for dry (black) and hydrated (blue) MIL-53(Ga) at 650~K. Bottom: free energy plot as a function of Ga--O and O--H distances (red is lower, blue is higher), with the bound and unbound states depicted. Figure adapted from Ref.~\citenum{Haigis2015}.}
    \label{fig:MIL53_metadynamics}
\end{figure}

Let us first focus on the modeling of hydrothermal stability. While it has been long known experimentally that many MOFs are unstable under exposure to water in liquid or vapor phase,\cite{Kusgens2009} computational descriptions of this lack of stability of frameworks have been difficult to achieve. The most-studied systems for this particular topic are the water-unstable MOFs of the IRMOF family, including the archetypical MOF-5 (also known as IRMOF-1). Initial studies focused mostly on the energetics of water adsorption and linker displacement, \cite{Low2009} and neglected thermal effects and entropic contributions. Later work by different groups highlighted the importance of hydration level (or water loading) and the role of collective effects in the breakdown \cite{DeToni2012, Bellarosa2012, Bellarosa2012b} by using first principles molecular dynamics to give microscopic insight into the mechanism behind the water instability. This showed the existence of a transient five-fold coordinated zinc species, which is stabilized by a hydrogen bond network with nearby physisorbed water molecules, leading at high enough vapor pressure to the linker displacement. Finally, a recent study of the hydrothermal stability of ``breathing'' MOF MIL-53(Ga) by Haigis et al.\cite{Haigis2015} combined the use of first principles molecular dynamics with a free energy method (namely metadynamics) to study the hydrothermal breakdown of the soft porous crystal --- as shown in Figure~\ref{fig:MIL53_metadynamics}. As MIL-53(Ga) is water-stable at room temperature, the use of metadynamics allowed the sampling of relatively high free energy barriers to show that the weak point of the structure is the bond between the metal center and the organic linker and elucidate the mechanism by which water lowers the activation free energy for the breakdown, and thus limits the thermal stability of MIL-53(Ga) in the hydrated phase.

A second example is that of computational studies of the mechanical and thermal stability of zeolitic imidazolate frameworks (ZIFs). Experimental studies of temperature-induced\cite{Bennett2011b} and pressure-induced amorphization\cite{Bennett2011, Cao2012} of ZIFs were originally reported in 2011, although the mechanism by which they occurred was unknown at the time. Later studies used DFT calculations of the ZIF structure and elastic properties\cite{Ortiz2012_PRL} as well as and Brillouin scattering on a ZIF-8 single crystal\cite{Tan2012} to show that this material has exceptionally low shear modulus, linked to soft motions involving the ZnN$_4$ tetrahedron. Later work, using classical molecular dynamics simulations at varying pressure, proposed a mechanism for the pressure-induced amorphization: the crystal-to-amorphous transition is triggered by the mechanical instability of ZIF-8 under compression, due to shear mode softening of the material.\cite{Ortiz2013} The occurrence of low shear moduli and subsequent sensitivity to pressure-induced amorphization were later shown, by both quantum and forcefield-based methods, to be generic features of ZIFs --- with pressure and temperature stability domains that depend on the topology of the individual frameworks.\cite{BouesselduBourg2014, Tan2015} Finally, we note that very recently, a computational study has been the first to address the issue of thermal stability and temperature-induced phase transitions of ZIFs, using first-principles molecular dynamics to investigate the melting of ZIF-4 and the structure and properties of the resulting MOF liquid.\cite{Gaillac2017}

\bigskip
\section{Perspectives}

Concluding this series of highlights of recent progresses in computational approaches to the physics and chemistry of soft nanoporous materials, we present here some avenues of research which offer, from our point of view, worthwhile perspectives for the future and address challenging open questions. First, we want to cite the increasing number of reports in which chemical reactivity plays a role in the properties of soft porous crystals, outside the more ``traditional'' role of heterogeneous catalysis. These phenomena, involving bond breaking and reformation and sometimes heavy reconstruction, require the precision of quantum chemistry methods but deal with complex large-scale systems, making their computational investigation difficult. As examples of such studies in the literature, we have already cited above examples of the hydrothermal breakdown of the MIL-53(Ga) framework at high temperature\cite{Haigis2015} and the crystal$\rightarrow$liquid melting transition in ZIF-4.\cite{Gaillac2017} Another recent example is the eye-catching case of the chemisorption of carbon dioxide in the Mg-based mmen-Mn2(dobpdc) metal--organic framework (mmen = $N,N'$-dimethylethylenediamine; dobpdc = $4,4'$-dioxidobiphenyl-$3,3'$-dicarboxylate).\cite{McDonald2015} There, the CO$_2$ molecules insert into metal-amine bonds by a cooperative process, inducing a reorganization of the amines into well-ordered chains of ammonium carbamate and providing large CO$_2$ separation capacities. This complex reaction was studied by a combination of \emph{in situ} infrared spectroscopy, \emph{ex situ} powder X-ray diffraction, periodic DFT calculations, \abinitio molecular dynamics, and XAS simulations, showcasing how a combination of several experimental and computational tools can yield microscopic insight.

In addition to that, a second trend observed is the need for multiscale modeling in complex systems. While the anomalous physical and chemical properties of soft porous crystals make them prime targets for applications such as adsorption separations, catalysis, drug delivery, and sensing, in in industrial application they would likely be used as nanostructured composites such as core--shell particles or mixed matrix membranes --- and not in the form of a powder of small crystals. There is thus a need for simulations of the behavior of composite materials, beyond the level of the perfect infinite crystal. Studies have shown, for example, the impact of crystalline particle size on the flexibility and adsorption properties of ZIF-8, where a decrease in the crystal size can lead to a suppression of adsorption-induced structural transitions.\cite{Zhang2014} This effect was attributed to a competition of adsorption behavior in the crystal bulk and at the external surface of the particle. Other studies have focused on the large-scale dynamics and mechanics of the flexible MOFs, looking beyond the single unit cell to the scale of the entire crystal or hybrid material. In 2012, Triguero et al.\cite{Triguero2012} proposed a mesoscale lattice-based model for the description of the adsorption-induced breathing of a single crystal of MIL-53, showing that three key physical parameters controlled the mechanism of the breathing at that scale: the transition energy barrier, the cell--cell elastic coupling, and the system size. At even larger scale, our group has developed the use of analytical models\cite{Coudert2016_Dalton} or numerical finite element methods modeling\cite{Evans2017} to predict the macroscopic mechanical response of composite microporous materials, where one of the components is a flexible material with anomalous properties, such as negative linear compressibility. We showed that the macroscopic response of the composite can deviate strongly from that of its individual components, and that the highly sought after mechanical properties of soft porous crystals may be tuned --- and sometimes completely overridden --- in a nanostructure composite.

Finally, we note that among the examples presented herein, there is a definite trend towards studies that combine both experimental and computational tools in order to provide insight into the mechanisms of materials flexibility. There is also an important focus on the dynamics of flexibility, through time-solved and space-resolved methods, as well as on the workings of these materials in real working conditions, obtained through \emph{in operando} studies. Together with the question of aging, these topics will likely prove to be key for both achieving better fundamental understanding, as well as using these novel highly flexible nanoporous materials materials in practical applications.

\bigskip
\section*{Acknowledgments}
We thank Anne Boutin and Alain H. Fuchs for stimulating discussions and continuing collaboration on the computational chemistry of flexible materials. We thank Jack D. Evans for critical reading and insightful comments.

\bibliography{article}
\bibliographystyle{rsc}

\end{document}